\documentstyle[11pt,fleqn]{article}
\newcommand{\mnref}[4]{\hangindent=0.5in \hangafter=1 #1
{\em #2 }{\bf #3}#4\par}
\newcommand{\apj}{Astrophys. J.,}
\newcommand{\mn}{Mon. Not. R. astr. Soc.,}

\def\gs{\mathrel{\raise1.16pt\hbox{$>$}\kern-7.0pt
\lower3.06pt\hbox{{$\scriptstyle \sim$}}}}
\def\ls{\mathrel{\raise1.16pt\hbox{$<$}\kern-7.0pt
\lower3.06pt\hbox{{$\scriptstyle \sim$}}}}
\def\um#1{$\mu$m\count255=0\let\a=.\let\aa=,\let\aaa=;\let\aaaa=:\if#1\a\count255=1\fi
\if#1\aa\count255=1\fi\if#1\aaa\count255=1\fi\if#1\aaaa\count255=1\fi\ifnum\count255=0
\ \fi#1}

\newfont{\emu}{cmu10}
\newfont{\dunh}{cmdunh10}

\def\w{{w($\theta$) }}
\def\ws{{w($\theta$)}}

\begin{document}

\begin{center}

\noindent{\Large{\bf The Edinburgh/Durham Southern Galaxy Catalogue - VI.\\
The Stability of w($\theta$)}}
\end{center}

\vspace*{0.2in}
\noindent
\renewcommand{\thefootnote}{{\fnsymbol{footnote}}}

\noindent{\bf R.C. Nichol} {\em Dept. of
Astronomy, University of Edinburgh,
Blackford Hill, Edinburgh EH9 3HJ}
$\!\!\footnote{\rm Present
address:  Department of Physics and Astronomy,
Northwestern University, Evanston, Illinois 60208, USA.}$

{\bf C.A. Collins} {\em Royal Observatory, Blackford Hill, Edinburgh EH9 3HJ}
$\!\footnote{\rm Present address: Physics Department,
Durham University, Durham, UK.}$

\vspace*{0.5in}
\parindent = 0pt

{\bf Abstract}

The form of the  galaxy angular correlation function derived from the
Edinburgh/Durham Southern Galaxy Catalogue (EDSGC)  has strong theoretical
consequences for models of structure formation. In this paper, we investigate
the effects of galactic extinction close to the south galactic pole and
plate--to--plate matching errors. These represent the most probable systematic
errors within the EDSGC which could affect the angular
correlation function. The distribution of extinction within the EDSGC area
was obtained using the Stark {\it et al.} HI
map and the IRAS $100\mu\rm{m}$ flux map. We find that the amplitude of
the whole EDSGC correlation function varies by less than ${\rm \delta
w}=0.003$ for reddening ratios $\rm{R_V}$ in the range $3.25\rightarrow6$. This
corresponds to a range in average extinction  of ${\rm
A_B}=0.12\rightarrow0.24$.  Simulations were carried out to assess the
contribution to the correlation function from correlated and uncorrelated plate
magnitude errors.  The only simulation that affected the large--scale power
seen
in \w was for correlated plate errors with a systematic plate--to--plate
offset of ${\rm \Delta m} =0.02$. This represented an overall $0.4$ magnitude
difference
between the ends of the EDSGC which was inconsistent with checks carried out
with external photometry. All other simulations had an insignificant effect  on
the large--scale form of the correlation function. These tests suggest that the
large--scale power seen in the EDSGC correlation function is due to intrinsic
clustering and is not an artifact of the construction or location of the
catalogue. A final corrected \w is presented at the end of this paper and
represents the best estimate of this function from the EDSGC.

\section{Introduction}

New observations of the two--point galaxy angular correlation function
(w($\theta$)) computed from three large--area automated galaxy catalogues
(Maddox {\it et al.} 1990, Picard 1991 and Collins, Nichol \& Lumsden
1992) have provided strong evidence for significant galaxy clustering
on large angular scales
($>5^{\circ}$) compared to the ``canonical'' function derived by Groth \&
Peebles
(1977) from the Lick galaxy
catalogue. The significance of these new
results is
that the popular
standard Cold Dark Matter (CDM) scenario of structure formation (Davis {\it et
al.} 1985), which could adequately
explain the Lick \w (Bond \& Couchman 1988), can not account
for the
level of galaxy clustering seen on large scales in these new galaxy catalogues.
Since all three independently constructed
catalogues find a similar amount of large--scale power in w($\theta$),
their combined correlation function has become
one of the most powerful
constraints on CDM and other theories of galaxy formation (see Wright {\it et
al.} 1992 and Efstathiou, Bond \& White 1992).
Due to the importance of this result it is vital to establish the robustness
of the large--scale power seen in \w and to demonstrate that it is really
due to intrinsic galaxy clustering.

This paper is concerned with the \w derived from the Edinburgh/Durham Southern
Galaxy Catalogue (EDSGC). This catalogue
was constructed from COSMOS measurements of 60 IIIaJ
UK Schmidt survey plates in a contiguous area of $1400\rm{deg}^2$, centered on
the South Galactic Pole (Heydon-Dumbleton, Collins \& MacGillivray 1989, paper
II). As mentioned above, Collins, Nichol \& Lumsden
(1992, paper III) presented the
galaxy angular correlation
function from the whole EDSGC and found
significant clustering of galaxies on angular scales
corresponding to $\simeq30\,h^{-1}\rm{Mpc}$.
In this paper,
we investigate the stability of that angular correlation function to
the effects of plate
matching errors
and variable
galactic extinction within the survey area.
These two effects are almost certainly the largest sources of
uncertainty within the catalogue and can, in principle,
introduce spurious clustering
signals which may mimic the large--scale power seen in $\rm{w}(\theta)$.

To provide
a measure of the size of the effect involved, consider a system of plates for
which there is an rms error in the background magnitude calibration of the
plates ($\Delta\rm{m}$). If the galaxy differential number counts take the form
$\rm{N}(\rm{m})=\rm{dex}(\rm{b}\rm{m})$, then from the definition
of $\rm{w}(\theta)$ as a variance estimator (see paper III)
we have
\begin{equation}
\delta{\rm{w}}\simeq(\frac{\delta{\rm{N}}}{\rm{N}})^2
\sim(2.3{\rm b}\Delta{\rm{m}})^2.
\end{equation}

For ${\rm b}=0.6$ (paper II), calibration errors between plates with an rms
of $\Delta\rm{m}=10\%$, give rise to $\delta\rm{w}\sim0.02$. This corresponds
to the amplitude of the EDSGC correlation function on a scale $\simeq 5^\circ$,
at a depth corresponding to $b_j\simeq18.5$. In view of the potentially large
affect systematic errors can have and the theoretical importance of the
large--scale power seen in w$(\theta)$, it is imperative to
fully demonstrate that plausible errors in the EDSGC do not affect the
conclusions drawn
from the observed galaxy correlation function.

The structure of this
paper is as follows. In Section 2 we correct the EDSGC for galactic
extinction using the Stark {\it et al.} (1992) HI column density maps of the
southern hemisphere.
A similar analysis is also carried out using the IRAS $100\,\mu\rm{m}$ fluxes
as described by Rowan--Robinson {\it et al.} (1991).
The resulting effect on the observed correlation function from
this analysis is discussed in this section. In Section 3 we simulate
both correlated and
uncorrelated plate--to--plate magnitude calibration errors and
discuss their affect
on the observed $\rm{w}(\theta)$.

\section{Galactic Extinction}

There has been considerable controversy over the level of extinction at the
SGP. The most widely assumed form for interstellar
extinction is the $\rm{cosec}\,|\rm{b}|$ extinction law. The
most famous example of such a model is that proposed
by Sandage (1973), which made the rather unrealistic assumption of canonical
dust-free
regions towards the galactic caps. Since the EDSGC is at high galactic
latitudes, $\rm{b}\leq-50^\circ$, the expected
extinction based on such an assumption is small. Another approach is to utilise
the correlation between the distribution of neutral atomic hydrogen and
dust. In a classic study,
Burstein \& Heiles (1978)
examined the relationship between HI column density, E(B--V) reddening and
the Lick galaxy counts. From an independent analysis of the same HI data,
Fisher \& Tully (1981)
concluded that there is $0.06$ magnitudes of extinction at the SGP.
More importantly, the HI data indicated significant patchy extinction across
most of the sky and demonstrated that the
smooth  $\rm{cosec}\,|\rm{b}|$ law was a simplistic description of galactic
extinction.

Recently, there have been two surveys carried out which have significantly
enhanced our knowledge of galactic extinction.
First, Stark {\it et al.} (1992) have recently completed a HI map covering
the sky
north of $\delta=-40^{\circ}$. This data was obtained using the Bell Laboratory
antenna with
a half-power beamwidth of $2.5^\circ$. Secondly, Rowan-Robinson {\it et al.}
(1991) have analysed the $100\,\mu\rm{m}$ intensity maps obtained from the IRAS
Zodiacal History File. This data represents the interstellar emission
from warm dust in the galaxy. Using a model for the interaction between dust
grains and the interstellar radiation field (Rowan-Robinson {\it et al.} 1990)
an
estimate of the dust column density can be made and the visual extinction
can be calculated from the intensity maps.

\subsection{Extinction Corrections}

The basic correlation between HI and visual extinction is given by

\begin{equation}
\rm{N(HI)} = \eta \rm{E(B-V)},
\end{equation}

where $\rm{N(HI)}$ is the hydrogen column density and $\eta$ is the
dust--to--gas ratio in suitable units. $\rm{E}(\rm{B}-\rm{V})$ is defined as

\begin{equation}
\rm{E}(\rm{B}-\rm{V})=\rm{A_{\rm B}}-\rm{A_{\rm V}},
\end{equation}

where $\rm{A_{\rm B}}$ and $\rm{A_{\rm V}}$ are the extinction estimates in
magnitudes
for the photometric
bands B and V respectively. It is convenient to relate the hydrogen column
density to the reddening ratio R, which is defined in a particular photometric
band (say V) as

\begin{equation}
\rm{R_{\rm V}}=\frac{\rm{A_v}}{\rm{E(B-V)}}.
\end{equation}

Then using equations 2, 3 and 4, we have

\begin{equation}
\rm{A_B}=\frac{\rm{N(HI)}}{\eta}(1+\rm{R_{\rm V}}).
\end{equation}

Using this formula,
the extinction estimates can be calculated from the HI
data assuming typical values
for $\rm{R_V}$ and $\eta$.
Unfortunately, the values of $\rm{R_V}$ and $\eta$
are highly uncertain and
vary by up to a factor of 2. However, typical values of these variables are
${\rm{R_V}}=3.5$ and
$\eta=5.2\times10^{21}{\rm cm^{-2}\,mag^{-1}}$ (Mathis 1990).
To illustrate the size of the extinction within the EDSGC
region, we note that the peak HI column density from the
Stark {\it et al.} data is $3.8 \times 10^{20}\,\rm{cm}^{-2}$ at $\alpha
\simeq0\,\rm{hr}$ and $\delta\simeq-40^\circ$. This corresponds to an
extinction of $\rm{A_{\rm B}}\simeq0.3$. At the SGP, the typical extinction
value is
$\rm{A_{\rm B}}\simeq0.1$, a result inconsistent with extinction free poles.

Using the IRAS $100\,\mu\rm{m}$ data and a model for the dust, Rowan--Robinson
{\it et al.} (1991) find a relation between $\rm{A_{\rm V}}$ and the
IRAS $100\,\mu\rm{m}$
intensity of the form

\begin{equation}
\rm{A_{\rm V}}=0.06\,\rm{I}(100)
\end{equation}

for $\ \mid\! b\!\mid>5^{\circ}\ $. At the SGP, $\rm{I}(100)=1.3\, \rm{MJy}\,
\rm{sr}^{-1}$ which, using equation 6, corresponds to $\rm{A_{\rm V}}=0.08$, a
value
consistent with that obtained from the HI estimate. Using $\rm{R_V}=3.5$, then
equation 6 can be converted to B band extinction and
the following formula is obtained,

\begin{equation}
\rm{A_{\rm B}}=0.08\,\rm{I}(100)\, .
\end{equation}

Using the formulae derived above,
three different extinction estimates were used to correct the EDSGC for
different values of $\rm{R_V}$ and $\eta$ given in the literature. These were:

\begin{enumerate}

\item[({\it i})]
Using the Stark {\it et al.} HI data with $\rm{R_V}=3.25$ and
$\eta=5.2\times10^{21}\,{\rm
cm^{-2}}\,{\rm mag^{-1}}$.
These values are close to typical estimates in the literature (Mathis 1990).

\item[({\it ii})]
Using the Stark {\it et al.} HI data with $\rm{R_V}=6$ and
$\eta=4\times10^{21}\,{\rm
cm^{-2}}\,{\rm mag^{-1}}$.
These represent the largest plausible values as quoted by Mathis (1990).
The peak extinction
then becomes $\rm{A_B}=0.6$, the mean extinction ${\rm A_B}=0.24$ and at the
SGP $\rm{A_B}=0.2$.

\item[({\it iii})]

Using the IRAS $100\, \mu \rm{m}$ data and equation 7 above.
\end{enumerate}

\subsection{The Effect on $\rm{w}(\theta)$ of the Extinction Corrections}

Extinction values for the 3 cases described were calculated
in the B band for each line--of--sight towards all the
galaxies within the EDSGC. The galaxy magnitudes were corrected simply by
subtracting the $\rm{A_B}$ values given by the equations above. We note here
that this procedure is not strictly correct, as we are using isophotal
magnitudes and extinction also effects the measured extents of the galaxies.
This means that the isophotal magnitudes are reduced by more than just
$\rm{A_B}$ (Phillipps, Ellis \@ Strong 1981). Our simplified extinction
correction can be justified on two counts. First, in this paper we are not
interested in calculating accurate extinction free photographic magnitudes.
Our primary concern is to determine the effect on the correlation function
of variations in the extinction on angular scales of a few degrees. Secondly,
the size of any diameter effect will be small. Using realistic galaxy profiles
we estimate that our magnitudes at 19.5 are within 0.1 of the total magnitude.
Corrections to the extinction estimates for the diameter effect will be
even smaller than this.  After correcting each galaxy for extinction, the
galaxies were then re--selected to a
depth of $b_j=19.5$ and the correlation function was calculated
over the identical area as was  used in paper III. A discussion of
the various $\rm{w}(\theta)$ estimators available is given in
paper III. For the correlation functions calculated here,
we use the estimator given by

\begin{equation}
1+\rm{w}(\theta)=\frac{n_{gg}}{n_{gr}}\frac{2\rm{N_r}}{\rm{N_g}},
\end{equation}

where $\rm n_{gg}$ is the number of discrete galaxy-galaxy pairs, $\rm n_{gr}$
is
the number of galaxy-random pairs, while $\rm{N_r}$ and $\rm{N_g}$ are the
total number
of random points and galaxies respectively.
The correlation functions for the 3 extinction corrections described above
are shown in Figure 1. The
$\rm{w}(\theta)$ estimates in this figure have been scaled to
the number density of the original uncorrected function given in paper III
using the scaling procedure described in that paper.
The number of galaxies found to the depth of $b_j=19.5$ for each trial were:
252021 for trial ({\it i}), 300994 for trial ({\it ii}) and
266820 for trial ({\it iii}).

Figure 1 indicates that the extinction corrections made very little difference
to the overall form of the correlation function. The
maximum deviation between the uncorrected \w at $b_j=19.5$ given in paper III
and a corrected \w
is $\delta{\rm w}\simeq0.003$ at $8^{\circ}$ (Figure 1). Using equation 1, this
translates to a 4\% rms fluctuation in the number counts on this scale, which
is comparable to the known plate--to--plate fluctuation discussed later in this
paper. The major
result of these extinction corrections is that the large--scale power in
$\rm{w}(\theta)$ still remains, even when the most severe extinction
corrections are used.

\section{Systematic Plate--to--Plate Errors}

\subsection{Object Classification and Magnitude Calibration in the EDSGC}

As discussed in Section 1, if the EDSGC
is to be used for any statistical
measurement of the galaxy distribution, great care must be taken to ensure that
the survey is homogeneous. There are two primary sources of systematic error
which dominate a digitised galaxy survey; star-galaxy classification and
plate magnitude calibration errors. Details of the image classification
algorithm used in constructing the EDSGC have been extensively discussed in
paper II.
In that paper we presented statistics
which indicated that the final catalogue is $>95\%$ complete with
$<10\%$ stellar contamination and at a depth corresponding to
$b_j\sim 19$, has only $\simeq 3\%$ rms residual
variation in the number of objects classified as galaxies. The image
classification was carried out for each Schmidt plate in turn using an
automated classification algorithm (Paper II).
The systematic variations in number density  between plates resulting
from mis--classification errors are therefore uncorrelated.

The EDSGC was
photometrically calibrated using galaxy CCD sequences on approximately
half of the
60 fields constituting the survey.
These sequences were used as "tie" photometric points to
which surrounding plates were calibrated using the $\simeq0.3^\circ$ overlap
region between the plates.
The histogram of magnitude offsets between fields in the EDSGC is shown in
Figure 1 of paper III and indicates that residual plate
calibration errors in the survey are $\simeq 5\%$ rms.
The density of galaxy calibration sequences for the EDSGC is
deliberately high to ensure that calibration errors are not
propagated across plate boundaries.
External checks with CCD photometry published by Maddox, Efstathiou \&
Sutherland (1990) and
Colless (1989) show no significant systematic variations either as a function
of
magnitude or position in the EDSGC (Nichol 1992).
Details concerning the photometric
calibration of the EDSGC and the external photometry checks
will be published in a
forthcoming paper.

However, since approximately one half of the plates in the EDSGC
are calibrated from adjoining fields which have CCD sequences, calibration
errors across the survey could, in principle, be correlated, thus leading to
underlying gradients across the EDSGC. This point has been made in
some detail for the APM galaxy survey by Fong, Hale--Sutton \& Shanks (1992).
In view of this, we have examined the effects of both correlated and
uncorrelated plate--to--plate number density
errors on the galaxy angular correlation function.

\subsection{Random Plate--to--Plate Variations}

In order to estimate the effect of random plate--to--plate errors on the
correlation function, simulations using randomly generated datasets
were carried out.
Random catalogues were generated over an identically shaped area to that used
in paper III for computing the EDSGC \w
and the number densities on each plate were varied by selecting
number densities from a Gaussian of
known width. In this way, the width of the Gaussian dictates the rms
percentage variation
in number density between the plates.
Figure 2 shows an example of one realization of these
simulations and
should help to illustrate the procedure used. The
plate boundaries are visible due to pixels lying on the plate boundaries having
no plate identification. Variations in the plate number density are clearly
visible.

Simulations were carried out for 10\%, 20\% and 30\% plate--to--plate number
density variations. For
each rms density variation, about 50 realizations were carried out, each
containing approximately 70,000 points binned as shown in Figure 2.
This number of points is close the number of observed galaxies within the
EDSGC \w
area at the Lick depth $b_j=18.6$ (68456 galaxies) and therefore, removes
the need for
re--scaling any of the correlation functions.
The \w for each realization
was calculated in the usual manner,
as described by equation 8, and were
averaged together to produce a single correlation function
(${\rm w_p (\theta)}$) for each of the different rms density fluctuations.
The individual correlation functions were not scaled to the same depth before
averaging because
the difference between their respective number densities was small
($1\sigma\simeq 0.9\%$).
The $\rm{w_p}(\theta)$ results
for the simulations at 10\%, 20\%, and 30\% are shown in Figure 3.
The error bars are $1\sigma$ derived from the scatter seen between the
individual correlation functions.

Following Geller, de Lapparent \& Kurtz (1984), the true or intrinsic
correlation function (${\rm w_{int}}$) can be recovered using the formula

\begin{equation}
{\rm w_{obs} = w_p w_{int} + w_p + w_{int},}
\end{equation}

where $\rm w_{obs}$ is the uncorrected or observed correlation function.
This equation assumes that the plate-to-plate errors and the intrinsic galaxy
distribution are uncorrelated. The intrinsic correlation functions for each of
the three plate density
variations above are
shown in Figure 4 along with the scaled uncorrected \w taken from paper III.

Figure 4 clearly shows that both the 10\% and 20\%
simulations have very little effect on the observed correlation function.
Only values above these
variations does it become significant. This would correspond
to an rms fluctuation in the
plate zero points of ${\rm \Delta m}\ge 0.15$, which is a factor of 3
larger than the
actual measured rms plate--to--plate variations for the EDSGC. In addition,
it is significantly greater than
the observed discrepancies with external photometric comparisons. However, the
most
striking feature of all the simulations is that their effect on the observed
correlation function is insignificant on scales greater than $4^{\circ}$.

\subsection{Correlated Plate--to--Plate Variations}

It is easy to envisage that correlated plate--to--plate variations in the
number density of objects on each plate can have a significant effect on the
observed correlation function. If the zero points of plates are systematically
under or over estimated from one side of the survey to the other, then in
principle, there can be a large discrepancy in limiting magnitude between the
two ends of the survey. For example, if there is a
systematic error in the zero points of adjacent places of ${\rm \Delta
m}=0.05$,
this would result in one side of the survey
being $0.6$ magnitudes fainter. However, this is too large an offset
not to have been
noticed and the external photometric ``ties'' prevents such a large offset,
but the effect
could, in principle, operate at a lower level.
Fong, Hale-Sutton \& Shanks (1992) have highlighted the
effects of correlated errors on
$\rm{w}(\theta)$ for the APM galaxy survey (Maddox {\it et al.} 1990)
and are able to remove the disagreement between the
APM $\rm{w}(\theta)$ and the Lick
$\rm{w}(\theta)$ by
simulating the effects of these small systematic variations. They claim that an
offset of ${\rm \Delta m}=0.01$ between plates would not have been detected by
the
checks the APM group carried out using external photometry, but is large enough
to dramatically change the form of the correlation function.

In order to investigate the stability of the EDSGC correlation function
to correlated zero point errors, we  simulated linear gradients across the
survey in addition to a 10\% random plate--to--plate component as described
above. Once again, these simulations were carried out using the identical area
and plate configuration as used in paper III.
Starting in the northeastern corner of the EDSGC
$\rm{w}(\theta)$
area (Field 532) the required ${\rm \Delta m}$ was added to adjacent plates so
that a
gradient was simulated in both declination and right ascension. Figure 2 shows
one realization of these simulations along with
the actual distribution of plate number densities in the EDSGC
$\rm{w}(\theta)$ area for comparison.
Simulations were carried out for ${\rm \Delta m}=0.01$, $0.015$ and $0.02$; in
each
case, over 50 realizations were used. To illustrate the magnitude of this
effect,
a gradient of ${\rm \Delta m}=0.02$ would
result in a global offset of 0.3 magnitudes in right ascension between the ends
of the EDSGC.

As detailed above, the correlation function (equation 8) was calculated for
each realization and then averaged together to produce a single ${\rm
w_p(\theta)}$ for each ${\rm \Delta m}$ gradient. Unlike the random
simulations,
the individual correlation functions were scaled to the same number density
(70000 points) before averaging because the differences in their densities was
significant. Figure 5 shows the three simulations along with a representation
of the $1\sigma$ error bars derived from the observed scatter between the
individual realizations. Figure 6 presents the effect of subtracting these
functions, using equation 9, from the uncorrected \w at the Lick depth.

The ${\rm\Delta m}=0.01$ ${\rm w_p(\theta)}$ has little effect on the overall
form of
the observed EDSGC correlation function. This is the systematic
magnitude gradient used
by Fong, Hale--Sutton \& Shanks (1992) in their analysis of the photometric
calibration of the APM survey. As already stated, they claim this gradient can
remove all the large--scale power (i.e. scales $>10^\circ$) from the APM \w.
This is clearly not the case
for the EDSGC \w, and in addition, the same is true for the ${\rm \Delta
m}=0.015$
gradient (Figure 6). However, for gradients as high as ${\rm \Delta m}=0.02$,
it is
possible to remove the large--scale power from the EDSGC \w. However, as
detailed above, this would require a 0.3 magnitude offset between the ends of
the survey in right ascension and a 0.1 magnitude offset in declination. These
values are inconsistent with both internal and external photometry checks
(Heydon--Dumbleton 1989, Nichol 1992, Collins \& Nichol, in preparation).

\section{The Best Estimate EDSGC \w}

To facilitate
accurate comparisons between observation and theory,
Figure 7 shows the best estimate of the EDSGC
${\rm w(\theta)}$ at a depth of $b_j=19.5$.
This function was calculated from the EDSGC after it had been
corrected for extinction using the Stark {\it et al.} data with the
most accepted values
of the reddening coefficient and of the dust--to--gas ratio, as given in
Section 3.1.
In addition, the averaged 10\% random
plate--to--plate correlation function, detailed in Section 4.2, has been
subtracted after appropriate scaling to the depth of
$b_j=19.5$. A fluctuation of this order is expected on the basis of the
rms plate--to--plate offsets given in paper III (0.05 magnitude uncertainty
per plate, equivalent to 7\% plate--to--plate fluctuations).
As discussed at length above, the subtraction of
this function makes little difference to the overall form of the
observed correlation function.
A linear plot of the
EDSGC \w is also shown which includes the observed errors on half the data
points.
These errors were calculated by
splitting the EDSGC into 3 independent areas as detailed in paper III.

\section{Discussion and Conclusions}

In Section 2, corrections made to the EDSGC
for galactic extinction were discussed.
It was clearly demonstrated that these corrections
had only a small effect on the overall
form of the observed correlation function.
In addition, even when the most extreme observed
values for the
reddening and the dust--to--gas ratio were used,
the large--scale power in \w remained.
These results are consistent with
the original motivation for constructing the catalogue at high galactic
latitudes around the South Galactic
Pole, thereby reducing galactic extinction.

The effect of systematic plate--to--plate errors on the observed \w were
discussed in Section 3. The findings of paper II were reviewed which showed
that object mis--classification varied by $\simeq 3\%$ between
plates within the EDSGC and that they were uncorrelated.
This effect was considered negligible compared to the
photometric calibration errors within the EDSGC.
Simulations carried out in this paper
show that plate--to--plate photometric calibration errors
make an insignificant contribution to the large--scale power
seen in the observed ${\rm w(\theta)}$,
especially if the errors are uncorrelated. Plate matching errors only become
significant if they are correlated across the whole survey area at a level of
${\rm \Delta m}\ge 0.02$ magnitudes between each plate. This would lead to an
overall 0.4
magnitude offset between the edges of the survey which is inconsistent with
external photometric checks. However, it should be noted that amount of
external photometry within the EDSGC area is limited and therefore, a
definitive check will have to wait until more data is available.

Plate magnitude calibration and object classification are certainly the
largest systematic effects which could corrupt a galaxy
catalogue such as the EDSGC. The checks presented here provide
strong evidence
that these errors do not affect the large--scale power
seen in the EDSGC \w and indicate that the EDSGC \w is robust and reliable.
Therefore, due to the high degree of scrutiny levelled at
the angular correlation function,
it must now be one of the most powerful observational constraints on theories
of
galaxy formation (see, for example, Wright {\it et al.} 1992).

\section{Acknowledgements}

Once again, we would like to thank the excellent support of the UK Schmidt
Unit and the COSMOS Unit at the Royal Observatory Edinburgh. We gratefully
acknowledge Andrew Connolly and Stuart Lumsden for stimulating discussion
during this project. We thank Mark Jones at QMW for supplying us with the IRAS
and Stark {\it et al.} data. Finally, RCN thanks
Mel Ulmer for his support during the final stages of this project and SERC
studentships for partial financial support. An electronic version of
all the data can be obtained from the authors via email (11340::NICHOL or
nichol@ossenu.astro.nwu.edu).

\section{References}

\mnref{Bond, J.R. \& Couchman, H., 1988.}{Proceedings of the Second Canadian
Conference in General Relativity and Relativistic Astrophysics.}{}{eds. Coly,
A. \& Dyer, C., World Scientific Press, Singapore.}
\mnref{Burstein, D. \& Heiles, C., 1978.}{\apj}{225}{, 40.}
\mnref{Colless, M., 1989.}{\mn}{237}{, 799.}
\mnref{Collins, C.A., Nichol, R.C. \& Lumsden, S.L., 1992.}{\mn}{254}{, 295
(Paper III).}
\mnref{Davis, M., Efstathiou, G., Frenk, C.S. \& White, S.D.M.,
1985.}{\apj}{292}{, 371.}
\mnref{Efstathiou, G., Bond. D. \& White, S.D.M., 1992.}{\mn}{}{ preprint.}
\mnref{Fisher, J.A. \& Tully, R.B., 1981.}{Astrophys. J. Lett.,}{243}{, L23.}
\mnref{Fong, R., Hale--Sutton, D. \& Shanks, T., 1992.}{\mn}{257}{, 650.}
\mnref{Geller, M.J., de Lapparent, V. \& Kurtz, M.J., 1984.}{Astrophys. J.
Lett.,}
      {287}{, L55.}
\mnref{Groth, E.J. \& Peebles, P.J.E., 1977.}{\apj}{217}{, 385.}
\mnref{Heydon-Dumbleton, N.H., Collins,C.A., \& MacGillivray,
      H.T., 1989.} {\mn}{238}{, 379 (Paper II).}
\mnref{Heydon-Dumbleton, N.H., 1989.} {Ph.D. Thesis.}{}{ University of
Edinburgh.}
\mnref{Maddox, S.J., Efstathiou, G., Sutherland, W. \& Loveday, J.,
1990.}{\mn}{242}{, 43p.}
\mnref{Maddox, S.J., Efstathiou, G. \& Sutherland, W., 1990.}{\mn}{246}{, 433.}
\mnref{Mathis, J.S., 1990.}{Ann. Rev. Astron. Astro.}{28}{, 37.}
\mnref{Nichol, R.C., 1992.}{Ph.D. Thesis.}{}{ University of Edinburgh.}
\mnref{Picard, A., 1991.}{Astrophys. J. Lett.,}{368}{, L7.}
\mnref{Phillipps, S., Ellis, R.S. \& Strong, W., 1981.}{\mn}{197}{, 1981.}
\mnref{Rowan--Robinson, M., Hughes, J., Leech, K., Vedi, K. \& Walker, D.,
1991.}{\mn}{249}{, 729.}
\mnref{Rowan--Robinson, M., Hughes, J., Vedi, K. \& Walker, D.,
1990.}{\mn}{246}{, 273.}
\mnref{Sandage, A., 1973.}{Astrophys. J.,}{183}{, 711.}
\mnref{Stark, A.A., Gammie, C.F., Wilson, R.W., Balley, J., Linke, R.A.,
Heiles, C. \& Hurwity, M., 1992.}{Astrophys. J. Suppl.,}{79}{, 77.}
\mnref{Wright {\it et al.}, 1992}{Astrophys. J. Lett.,}{396}{, L13.}

\newpage

\section{Figure Captions}

{\bf Figure 1: } A comparison between the uncorrected full 60--plate
w$(\theta)$ at $b_j=19.5$ from paper III, and the 3 different
extinction corrected correlation functions described in the text.
All the functions have been scaled to the number density of the original
w$(\theta)$ in paper III.
Only every
third point has been plotted to avoid overcrowding.

\vspace*{0.2in}

{\bf Figure 2: } Specific examples of the plate--to--plate simulations
described in this paper. The plots are a true sky
projection of all the plates in the area used to compute \ws.
The top plot shows the distribution of actual plate
number densities within the EDSGC, the middle one is an example of the 10\%
random plate--to--plate simulations and the bottom plot is an example of the
$\Delta m=0.015$ correlated plate--to--plate simulations.
Pixels on the boundaries of plates were flagged as shown and were not used in
the calculation of \ws.
Drill holes around bright nearby stars were also included in the simulations.

\vspace*{0.2in}

{\bf Figure3: }This figure shows the averaged 10\%, 20\% and 30\% correlation
functions for the 3 sets of random plate--to--plate simulations carried out.
The error bars are $1\sigma$.

\vspace*{0.2in}

{\bf Figure4: } The effect of subtracting the averaged random plate--to--plate
simulations in Figure 3 from the observed \w at $b_j=18.6$ taken from paper
III.

\vspace*{0.2in}

{\bf Figure 5: } This figure displays the averaged correlation
functions for the correlated plate--to--plate simulations with $\Delta m=0.01,
0.015\, \&\, 0.02$. One $\sigma$ error bars have again been presented.

\vspace*{0.2in}

{\bf Figure 6:} The effect of subtracting the correlated plate--to--plate
simulations in Figure 6 from the observed \w at $b_j=18.6$ from paper III.

\vspace*{0.2in}

{\bf Figure 7:} The best estimate of the EDSGC \ws. This is the observed \w
from paper III after it has been corrected for extinction and residual
plate--to--plate matching errors. Inset is a linear plot of the function
with every other error bar on the data points plotted.
These errors were derived from the observed scatter
in \w when the EDSGC was split into 3 separate areas.

\end{document}